\documentclass{article}
\usepackage{ijcai19}

\usepackage{times}
\usepackage{soul}
\usepackage{url}
\usepackage[hidelinks]{hyperref}
\usepackage[utf8]{inputenc}
\usepackage[small]{caption}
\usepackage{graphicx}
\usepackage{amsmath}
\usepackage{booktabs}
\begin{document}

\title{Data, Power and Bias in Artificial Intelligence}

\author{
Susan Leavy$^{1,2}$ \and
Barry O'Sullivan$^{1,3}$ \And
Eugenia Siapera$^2$ \\
\affiliations
$^1$Insight Centre for Data Analytics\\
$^2$School of Information and Communication Studies, University College Dublin, Ireland\\
$^3$School of Computer Science and Information Technology, University College Cork, Ireland\\
\emails
\ susan.leavy@ucd.ie$|$b.osullivan@cs.ucc.ie$|$eugenia.siapera@ucd.ie
}


\maketitle
\begin{abstract}
Artificial Intelligence has the potential to exacerbate societal bias and set back decades of advances in equal rights and civil liberty.
  Data used to train machine learning algorithms may capture social
   injustices, inequality or discriminatory attitudes that may be
   learned and perpetuated in society. 
  Attempts to address this issue are rapidly emerging from different
   perspectives involving technical solutions, social justice and data
   governance measures.
  While each of these approaches are essential to the development of a comprehensive solution, often discourse associated with each seems disparate.
  This paper reviews ongoing work to ensure data justice, fairness
   and bias mitigation in AI systems from different domains exploring
   the interrelated dynamics of each and examining whether
   the inevitability of bias in AI training data may in fact be used
   for social good. 
  We highlight the complexity associated with 
   defining policies for dealing with bias.
  We also consider technical challenges in
   addressing issues of societal bias.
\end{abstract}

\section{Introduction}

The detrimental effects of unfairness and bias in artificial intelligence (AI) on equal rights in society have been well documented. Predictive policing systems trained on historical police records are being increasingly used to forecast criminal behaviour even though the accuracy of this data has been called into question~\cite{richardson2019dirty}. Racial bias was uncovered in an algorithm used to support healthcare decisions \cite{obermeyer2019dissecting}. While bio-metric technology such as facial recognition is being increasingly integrated into core security and border control infrastructures, their accuracy is lowest on darker-skinned females~\cite{buolamwini2018gender}.  In search engines and recommender systems, race and gender discrimination  was uncovered across multiple online platforms~\cite{datta2015automated,lambrecht2019algorithmic,noble2018algorithms}. In each of these cases the effects of bias in AI was to discriminate against those already marginalised in society.

Bias and discrimination in AI follows a long history of injustice resulting from the way people are counted, represented and classified in data~\cite{perez2019invisible,hickman1997devil}. To those affected by injustice, whether unfair decisions are made by human or machine is not the most pressing issue. It is perhaps for this reason that the focus of analysis in terms of social justice, critical race studies and feminism is not solely restricted to artificial intelligence but algorithms, socio-technical systems and automated decision making.

In parallel, within the field of AI, approaches to ensuring fairness and justice often assume that entirely new theoretical and ethical frameworks are required. However, as is clear in relation to the use of historical police records to predict future crime~\cite{lum2016predict}, the same critical perspectives and activism that brought social justice in the past are fundamental to preventing bias and discrimination in AI. This paper explores this issue examining how research grounded in social justice, critical race studies and feminism  could form the foundation of a new approach to data collection and curation for AI. 
We also consider technical challenges in
 addressing issues of societal bias.

\section{The Myth of Objectivity}

Approaches to the prevention of bias and discrimination in AI have included developing fairness-aware machine learning algorithms, addressing bias in data and modifying learned models. The central aim of this work is to assume the status of objectivity and neutrality. However,  there is increasingly widespread acknowledgement of the impossibility of objectivity in data-driven AI~\cite{meredith2018artificial,o2016weapons} and of bias as ``\emph{an unavoidable characteristic of data collected from human processes~''}~\cite{dignum2019responsible}. This sentiment was aptly captured by Bartoletti~\shortcite{ivana} in her assertion that 
\emph{``it is time we acknowledge that data is simply not neutral, and, as such, every single decision and action around data is a political one.}''  This acknowledgement of the inevitability of bias necessitates a reform of the process of collating and curating training data for machine learning, incorporating perspectives across multiple disciplines.

Bringing to light the extent and complexity of damaging constructs of race and gender and how it is embedded in both contemporary and historical data makes it clear how unavoidable it is that data reflects some form of bias or captures the effects of social injustice.  Noble~\shortcite{noble2018algorithms} uncovered a series of cases where data-driven racial profiling of individuals resulted in the repetition of historical injustices against people of colour through what she termed \emph{`technologically redlining'.} How \emph{`runaway feedback loops'} are generated by machine learning algorithms when they are trained on data large datasets that capture the results of a history social injustice and how this exacerbates existing discriminatory practices in society is detailed by Gebru~\shortcite{gebru2019oxford}. 

The process of data collection and curation for machine learning algorithms, from the standpoint of feminism or critical race studies, can never be objective. Every dataset that aims to represent humans and their behaviour does so in accordance with a world-view or ideology that may be assimilated into an AI system. Data collection and curation for machine learning algorithms is therefore transformed into a political act of curating a world-view that one intends to perpetuate through an AI system. 




\section{Beyond Bias}

In contrast to the increasingly widespread acknowledgement of the impossibility of objectivity in data, most current technical approaches to bias in training data for machine learning aim to re-balance data and render it neutral in relation to protected classes such as race or gender. These include methods for data augmentation, re-sampling, re-weighting, swapping labels and removing dependencies between characteristics~\cite{kamiran2012data,feldman2015certifying}.  

In relation to language based datasets, methods to uncover and address bias include the use of word embedding models trained on text to uncover social and intersectional bias~\cite{tan2019assessing}. Stereotypical associations in text were amended by disassociating relationships between entities in the trained models~\cite{bolukbasi2016man} and by swapping the gender of entities in text~\cite{zhao2018gender,park2018reducing}. However, the evaluation of bias in these studies focus on specific aspects such as stereotypical associations of race and gender in relation to employment or use implicit association tests as evaluation frameworks for bias in the study by Caliskan~\shortcite{caliskan2017semantics}. 

Within narrowly defined contexts, approaches to modifying data have been shown to reduce bias in algorithms. However, grounding the work in critical studies of race, feminist theories and social justice may serve to enrich the frameworks for evaluating fairness in how groups are represented in data. Furthermore, incorporating such perspectives diverts efforts away from a search for bias towards a more comprehensive interrogation of the power structures and ideologies of gender, race and social class that may be captured in the data. 

In assessing the ideology underlying a dataset, the critical framework proposed by  D'Ignazio and Klein~\shortcite{d2020data} presents a comprehensive approach to the evaluation of how gender is  represented in data, grounded in intersectional feminist thought. The authors develop feminist principles for data collection asserting that \emph{``data are not neutral or objective.''} This approach if embedded into the data collection process for machine learning would highlight potentially problematic patterns in existing data that may lead to discriminatory AI systems. It may also work towards the transformation of the process to one involving the curation of data collections that intentionally reflect an intersectional feminist representation of gender.  

By bridging critical theories of race, social justice or feminist theory with training machine learning algorithms in this way,  less focus may be placed on attempting to `fix' specific issues of bias in datasets that contain embedded problematic ideologies. Rather, a dataset may be designed and selected in order to capture a representation of the kind of concepts of race, gender and social class that should be perpetuated in an AI algorithm.


\section{Power in Data}

One response to bias in data resulting from an under-representation of particular groups, is to collect more data from that group of people. This has led to questions of unethical data gathering practices with the goal of balancing datasets~\cite{hawkins2018beijing}. Benjamin~\shortcite{benjamin2019race} examines the complex socio-political ramifications of this in the context of race, questioning \emph{``what does it mean to be included, and hence more accurately identifiable, in an unjust set of social relations?.''} This historical legacy of inequitable treatment of the poorer in society in relation to data collection is described by Eubanks~\shortcite{eubanks2018automating} as what she termed the \emph{`Digital Poorhouse.'} In attempting to balance data sets for training machine learning algorithms, it is crucial then to consider the social and political effects of data collection for different groups in society.


The central importance of fair data in preventing discrimination in  AI, highlights the central role played by curators of data for machine learning algorithms. The cognitive labour of classifying and labelling AI training data is increasingly conducted by underpaid women of colour from the global south~\cite{crawford2018anatomy,d2020data}. However, with the increased recognition that bias and discrimination in AI emanates largely from data, along with the deconstruction of the myth of algorithmic objectivity, it should follow that the human labour and value involved in data curation for AI will become more visible. It also emphasises the case for what Benjamin~\shortcite{benjamin2019race} describes as the \emph{`democratization of data'} through the inclusion of all perspectives, especially from those most vulnerable to discrimination,  into the design of technology.

\section{Governing Artificial Intelligence}

The protection of autonomy and the right to self-determination of every individual is central to research and activism at the intersection of technology and social justice.  By virtue of its capacity to learn, AI raises foundational ethical questions in relation to human autonomy and introduces new dimensions in terms of accountability for discrimination. Ensuring humans are accountable for decisions made by AI systems is key to ethical AI~\cite{o2016weapons}. To reflect the foundational nature of such questions, governing bodies are developing new frameworks for ethical, trustworthy and responsible AI primarily grounded in a human rights perspective~\cite{hleg2019high,oecd,us}. However, given the accelerating pace of the influence of AI in society there is also an urgency on the implementation of measures to prevent discrimination. 

In addressing the urgent need for governance, organisations are examining the use of existing data governance policies in the context of the collection and curation of machine learning training data~\cite{8610013,iso}.  There are also calls for standards on the kind of inferences that AI algorithms can make based on an individual's data \cite{wachter2019right}. However, the implementability of data governance regulations to ensure bias-free data remains a challenge. The central aim of ensuring bias-free data contrasts with the increasingly accepted position that data, particularly in relation to human behaviour,  is inevitably biased. Additionally, within AI literature there is a range of different definitions of fairness and methods of representing the concept mathematically~\cite{mehrabi2019survey}. Departing from the concept of bias, which implies the possibility of bias-free data and grounding data regulation instead in concepts of discrimination derived from well established legal definitions, may contribute towards unifying research across the disciplines of AI, policy and social justice.


\section{Technical Challenges and Bias}

 

A considerable challenge in dealing with bias in AI
 generally, and in machine learning in particular, is that
 there can be different definition of bias or what it
 means to be fair.
We consider here the often discussed COMPAS system for
 predicting recidivism.\footnote{\url{https://www.propublica.org/article/how-we-analyzed-the-compas-recidivism-algorithm}}
 Tables~\ref{tab-black} and~\ref{tab-white} are based on
  an analysis presented by Sumpter~\shortcite{Sumpter2018}
  that considered whether or not COMPAS exhibits racial bias.
 The answer is that it depends on what one defines as bias.
 
Kleinberg et al.~\shortcite{DBLP:conf/innovations/KleinbergMR17}
 considered three different fairness conditions that
 capture what is means for an algorithm to be
 fair and avoid exhibiting bias.
The first of these is that the algorithm should be
 \emph{well calibrated}.
Essentially, this means that the proportion of people
 that are positive in a population should be equal to the
 proportion of people that are positive in each subgroup
 of the population.
Comparing Tables~\ref{tab-black} and~\ref{tab-white} we
 see the that $1369/2174 = 63\%$ of African American
 defendants classified as high risk, while 
 $505/854 = 59.1\%$ of White defendants were classified
 as such.
This data is well-calibrated, suggesting the system does not
 exhibit a racial bias, since the proportion of high-risk people
 in each population is predicted to be roughly equal.
 
The second (and third) condition relates to balancing
 for the positive (resp. negative) class.
If this condition were to be violated it would mean that
 the likelihood that a positive (resp. negative) instance
 in one population
 is more likely to be identified than in the other.
For example, as Sumpter argues,
 Tables~\ref{tab-black} and~\ref{tab-white}
 show that $2174/3615 = 60\%$ of African Americans in COMPAS
 are considered higher risk, while this only $854/2464 = 35\%$
 for Whites; note that $1901/3615 = 53\%$ of African Americans reoffended while this was $966/2454 =39\%$ for White, suggesting
 that African Americans actually reoffended less than predicted
 while the opposite is true for Whites.
This suggests a strong racial bias.
 
Considering the mistakes that COMPAS makes for each racial group,
 Sumpter goes on to show that $805/1714 = 47\%$ of African
 Americans were
 wrongly predicted to reoffend, as compared with only 
 $349/1488 = 24\%$ of Whites.
On the other hand, only $532/1901 = 28\%$ of African Americans
 who reoffended were wrongly predicted as being lower risk,
 as compared with $461/966 = 48\%$ of Whites.
Again, this suggests a strong racial bias.

In other words, while COMPAS satisfies one of Kleinberg et al.'s
 definition of fairness (calibration), is fails the second and
 third.
This is not unusual.
In fact, Kleinberg et al. rigorously prove that these three 
 fairness conditions
 are incompatible except in very rare situations.
This implies
 that one is usually faced with a choice of which bias must be
 traded-off against another.
Eliminating bias from AI systems is often provably impossible.
 
\begin{table}
\caption{Recidivism rates in the COMPAS system for African American defendants (via [Sumpter, 2018]).}
\label{tab-black}
    \small
    \centering
    \begin{tabular}{l|rr|r}
    \hline
         African American &  High Risk & Low Risk & Total\\
    \hline \hline
         Reoffended & 1,369 & 532 & 1,901 \\
         Didn't reoffend & 805 & 990 & 1,714 \\
         \hline
         Total & 2,174 & 1,522 & 3,615 \\
    \hline \hline
    \end{tabular}
\end{table}

\begin{table}
    \caption{Recidivism rates in the COMPAS system
    for White defendents (via [Sumpter, 2018]).} 
    \label{tab-white}
    \small
    \centering
    \begin{tabular}{l|rr|r}
    \hline
         White &  High Risk & Low Risk & Total\\
    \hline \hline
         Reoffended & 505 & 461 & 966 \\
         Didn't reoffend & 349 & 1,139 & 1,488 \\
         \hline
         Total & 854 & 1,600 & 2,454 \\
    \hline \hline
    \end{tabular}
\end{table}

The challenges in dealing with bias are complex.
To address these issues, in addition to the many point made
 earlier, we must develop a set of
 AI methods that reason about bias as systems are being developed.
It might be helpful to think of bias-aware
 learning as an optimisation problem in which
 the properties of the learned concept and the acceptable
 tradeoff between fairness criteria could be be imposed
 as constraints on the learning
 process~\cite{DBLP:conf/cp/BessiereHO09}.
 
\section{Conclusion}


In considering the challenges of bias in AI, it is important
 to build upon existing principles and frameworks, e.g.
 critical intersectional feminist and critical race theories.
Dealing with bias in an objective way is very challenging
 and the possibility of unbiased data is not a
 sufficient condition for unbiased AI.
In reasoning about bias and fairness is is critically important
 to involve those who might be directly impacted by AI system
 in the process of designing, curating, testing these systems,
 and especially in the `deconstruction' of language-based data.
It is important that we interrogate notions of accuracy
 that may lead to more
 surveillance and data gathering on marginalised groups, and
 the role of such marginalised groups in data curation is important
 to acknowledge.
 
\paragraph{Acknowledgement.}
This publication has emanated from research conducted with the financial support of Science Foundation Ireland under Grant number 12/RC/2289-P2  which is co-funded under the European Regional Development Fund.
 
\bibliographystyle{named}
\bibliography{ijcai19}

\end{document}